\documentstyle[12pt,epsfig,rotate]{article} 
\newlength{\dinwidth}                       
\newlength{\dinmargin}                      
\def\lsim{\mathrel{\rlap{\lower4pt\hbox{\hskip1pt$\sim$}}
    \raise1pt\hbox{$<$}}}                
\def\gsim{\mathrel{\rlap{\lower4pt\hbox{\hskip1pt$\sim$}}
    \raise1pt\hbox{$>$}}}                

\newcommand{\xpom}{x_{_{I\!\!P}}}
\newcommand{\gapprox}{\stackrel{>}{_{\sim}}}
\newcommand{\lapprox}{\stackrel{<}{_{\sim}}}
\newcommand{\pom}{I\!\!P}

\newcommand{\scaption}[1]{\caption{\protect{\footnotesize  #1}}}

\newcommand{\slowpi}{\pi_{\mathit{slow}}}
\newcommand{\gevsq}{\mathrm{GeV}^2}

\newcommand{\diiii}{\frac{{\rm d}^4\sigma_{ep\rightarrow eXY}}
  {{\rm d}x\,{\rm d}Q^2\,{\rm d}\xpom\,{\rm d}t}}

\def\Rd{R^D}
\begin{document}
\vspace*{1cm}
\begin{center}  \begin{Large} \begin{bf}
Future Diffractive Structure Function \\
  Measurements at HERA\footnote{To appear 
in the proceedings of the workshop ``Future Physics at HERA'',
DESY, Hamburg, 1996.}\end{bf}  \end{Large} \\
  \vspace*{5mm}
  \begin{large}
A.~Mehta$^a$, J.~Phillips$^b$, B.~Waugh$^c$ \\
  \end{large}
\end{center}
$^a$ Rutherford-Appleton Laboratory\\
$^b$ University of Liverpool\\
$^c$ University of Manchester\\
\begin{quotation}
  \noindent {\bf Abstract:} The purposes and possibilities of future
  diffractive structure function measurements at HERA are presented.
  A review of the current range and accuracy of the measurement of
  $F_2^{D(3)}(\beta, \xpom, Q^2)$ is presented and an estimate of the
  precision of future measurements is given.  A feasibility study is
  performed on the measurement of the structure functions
  $F_2^{D(4)}(\beta, \xpom, Q^2, t)$, $F_{2~{\it charm}}^{D}$,
  $R^{D(3)}$, $R^{D(4)}$ and $F_L^{\pom}$. Included in this study are
  estimates of the integrated luminosity required, the analysis
  techniques to be employed and values of systematic error that could
  be expected.

\end{quotation}
\section{Introduction}
The first pioneering measurements of the deep inelastic diffractive
cross section at HERA have yielded great insight into the mechanism of
diffractive exchange and the partonic structure of the
pomeron~\cite{H1F2D93,ZEUS_MASS93,H1F2D94,H1WARSAW}. 
The precision of present measurements is, however, inadequate for the 
study of many quantitative aspects of this picture. The origin of the gluonic
structure of the pomeron and the interface between soft and hard
physics are unresolved questions that would benefit from the higher
luminosity offered by HERA in the years to come. In addition, the 
substantially different partonic structure of the 
pomeron in comparison to usual hadrons makes diffraction
a challenging test for perturbative QCD~\cite{MCDERMOTT}.
Furthermore, it has been suggested that the emerging data from HERA 
may provide fundamental insight into the non-perturbative aspects 
of QCD, leading to a complete understanding of 
hadronic interactions in terms of a Reggeon Field Theory~\cite{ALAN}.

In this paper we study the measurements that can be made of inclusive
diffractive cross sections with the luminosities which may be achieved
in the future running of HERA. We evaluate what precision is possible
for these measurements, and attempt to highlight those measurements
which will be of particular relevance to achieving significant
progress in our understanding of QCD .


\section{Experimental Signatures of Diffractive Dynamics}
 A prerequisite for any precision measurement of a hadronic cross section
 is that it must be defined purely in terms of physically observable 
 properties of the outgoing particles. Since there is, as yet, only the
 beginnings of a 
 microscopic understanding of the mechanism responsible for diffractive 
 dynamics, and since there are a plethora of wildly different predictions for 
 how the diffractive cross section should depend on all of the kinematic 
 variables, 
 it is the experimentalist's job to provide a well defined measurement 
 unfettered by any {\em ad hoc} assumptions about how the cross section
 should behave. 

 For the interaction {\mbox{$ep\rightarrow eX$ ({\em $X=$Anything})}} 
 the cross section is usually measured differentially as a function 
 of the two kinematic variable $x$ and $Q^2$ defined as follows:
\begin{equation}
x=\frac{-q^2}{2P\cdot q},
\,\,\,\,\,\,\,\,\,\,\,\,\,\,\,\,\,\,\,\,\,\,\,\,\,\,\,\,
Q^2=-q^2,
\end{equation}
 where $q$ and $P$ are the 4-momenta of the exchanged virtual boson and 
 incident proton respectively, and $y$ is the fractional energy loss
 of the electron in the rest frame of the proton. 
 The differential cross section
 $\frac{{\rm d}^2\sigma}{{\rm d}x{\rm d}Q^2}$ may then be 
 related to the proton structure functions $F_2$ and 
 $F_L=F_2\cdot R/(1+R)$ by:
\begin{equation}
\frac{{\rm d}^2\sigma}{{\rm d}x{\rm d}Q^2}=
\frac{2\pi\alpha_{em}^2}{xQ^4}
\left[ 2(1-y) + \frac{y^2}{1+R}\right] \cdot F_2(x,Q^2).
\end{equation}

\begin{figure}[htb]
 \begin{center}
   \epsfig{file=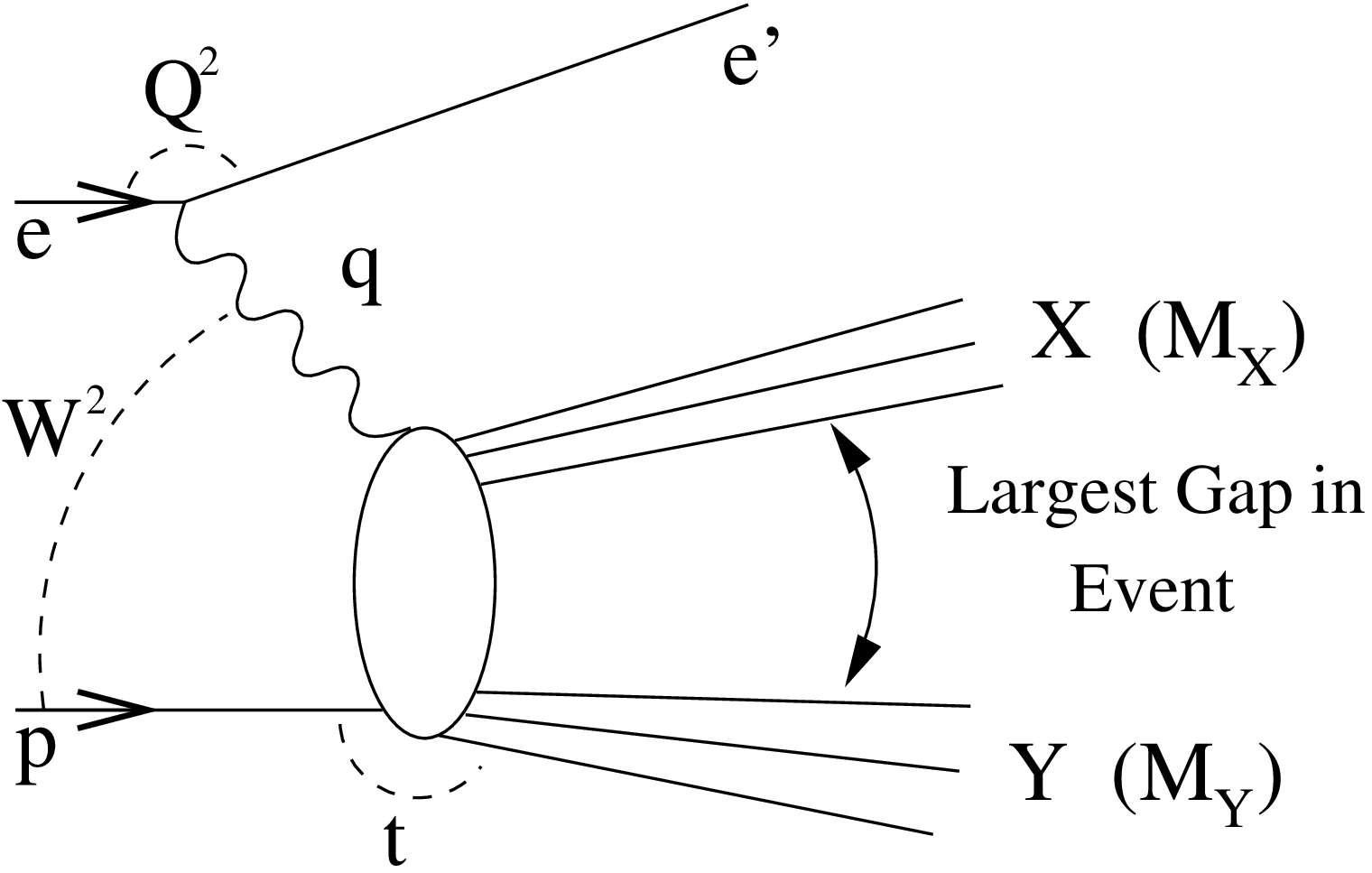,angle=270,width=0.4\textwidth}
 \end{center}
 \scaption {Schematic illustration of the method of selecting events
   to define a diffractive cross-section, showing the definitions of
   the kinematic variables discussed in the text. The systems $X$ and
   $Y$ are separated by the largest gap in true rapidity in the
   $\gamma^*$-$p$ centre of mass system.  The system $Y$ is nearest to
   the direction of the incident proton in this frame of reference.}
  \label{fig:diffXY}
\end{figure}

 Such an interaction may be further characterised by dividing the total
 hadronic final state (i.e. the final state system excluding the
 scattered lepton) into two systems, $X$ and $Y$, separated by the
 largest gap in true rapidity distribution of particles in the
 $\gamma^*$-$p$ centre of mass system, as shown in
 figure \ref{fig:diffXY}. 
 When the masses of these two systems, $M_X$ and $M_Y$, are both much
 smaller than the $\gamma^*p$ centre of mass energy $W$ then a large
 rapidity gap between the two systems is kinematically inevitable, and
 the interactions are likely to be dominantly diffractive.

  The two additional kinematic degrees of freedom may be specified 
  by introducing the variables $t$, $\beta$ and $\xpom$:
\begin{equation}
  t = (P-Y)^2,
\end{equation}
\begin{equation}
\xpom = \frac{Q^2+M_X^2-t}{Q^2+W^2-M_p^2} \simeq \frac{Q^2+M_X^2}{Q^2}\cdot x,
\end{equation}
\begin{equation}
\beta = \frac{Q^2}{Q^2+M_X^2-t} \simeq \frac{Q^2}{Q^2+M_X^2},
\end{equation}
 The hadronic cross section to be measured is therefore $\diiii$.  The
 presence of a rapidity gap may then be identified experimentally by
 use of the calorimeters and forward detectors of the experiments. For
 sufficiently small values of $M_X$, the system $X$ will be completely
 contained in the central calorimeters, allowing an accurate
 determination of $\beta$ and $\xpom$. Tagging of energy in the very
 forward direction allows both the mass of system $Y$ and the magnitude
 of $|t|$ to be constrained, but does not allow a direct measurement of
 $t$.

 A leading proton spectrometer (LPS) ``tags'' particles which follow
 closely the path of the outgoing proton beam separated by only a very
 small angle.  By tagging the particle at several points along it's
 trajectory and with detailed knowledge of the magnetic field optics
 between the interaction point and the tagging devices, it is possible
 to reconstruct the original vector momentum of the particle. Thus it
 is possible to select a sample of events containing a ``leading''
 proton with $E_p\sim E_p^{beam}$. An LPS permits an accurate
 measurement of $t$, and provides an independent method of selecting
 interactions sensitive to diffractive dynamics. Since there is no
 need for all the hadrons of system $X$ to be contained within the
 main calorimeter it is possible to make measurements at higher values
 of $\xpom$, so that subleading contributions to the cross section
 may be confirmed and further investigated.

 Regge phenomenology has been highly successful in correlating many 
 features of the data for both elastic and inelastic hadron--hadron  
 cross sections~\cite{ReggeFits}. At high energies (low $\xpom$) the 
 data are dominated by the contribution from the leading trajectory 
 of intercept $\alpha(0)\gapprox 1$. The most simple prediction of 
 Regge phenomenology is that at fixed $M_X$  the 
 cross section should vary $\propto \xpom^{-n}$ where $n$ is related
 to the intercept of the leading trajectory by $n=2\alpha(t)-1$.\footnote{
It is worth pointing out that at $Q^2>0$ then $\xpom$ is the basic 
 Regge variable, not $W^2$ as is sometimes assumed.} The size of 
 the rapidity gap is kinematically related to the value of $\xpom$ 
 by \mbox{$\Delta \eta \sim -{\rm ln} \xpom$}~\cite{XPDETA}. 
 It is easy to show that at fixed $M_X$ the rapidity gap distribution 
 is then related to $\alpha(t)$ by:
\begin{equation}
\frac{{\rm d}N}{{\rm d}\Delta\eta} \propto (e^{-\Delta\eta})^{2-2\alpha(t)}.
\end{equation}
Hence for $\alpha(t)<1$ the production of large rapidity gaps is
exponentially suppressed, whilst for $\alpha(t)>1$ it is exponentially
enhanced~\cite{Bjorken}. With $\xpom$ fixed, then extensions of the
simple Regge model are necessary to predict the $M_X$ dependence.  For
sufficiently large masses, it is perhaps possible to use M\"{u}ller's
extension of the optical theorem~\cite{Mueller} (``Triple Regge'') to
achieve this aim, and for small masses the vector meson dominance
model~\cite{VMD} may be appropriate.
 
 There is much evidence from hadron--hadron data~\cite{ReggeFits}, 
 and some from $ep$ data~\cite{H1WARSAW} that additional 
 (``sub--leading'') trajectories are necessary to reproduce the 
 measured hadronic cross sections. The Regge formalism permits $t$ 
 dependent interference between some of the different contributions 
 rendering problematic any theoretical predictions for the 
 dependence of the measured hadronic cross sections on $M_X$, 
 $x_{I\!\!P}$, $W^2$ or $t$.

 Estimates of the diffractive contribution to both deep--inelastic
 and photoproduction cross sections have been made using the 
 $M_X$ dependence at fixed $W$~\cite{ZEUS_MASS93}. 


\section{Future Measurements of {\boldmath $F_2^{D(3)}$}}
 The 1994 H1 data \cite{H1F2D94,H1WARSAW}, corresponding to 
 $\sim 2$~${\rm pb^{-1}}$, have allowed a measurement of $F_2^{D(3)}$ to 
 be made in the kinematic region $2.5 < Q^2 < 65 $ GeV${}^2$, 
 $0.01 < \beta <0.9$ and $0.0001 < x_{I\!\!P} < 0.05$. 
 For $Q^2<8.5\,{\rm GeV^2}$ this
 was achieved by taking data with the nominal interaction point shifted
 by $+70\,{\rm cm}$ allowing lower angles to be covered by the backward
 electromagnetic calorimeter.

The dependence of $F_2^{D(3)}$ on $x_{I\!\!P}$ was found not to depend
on $Q^2$ but to depend on $\beta$, demonstrating that the
factorisation of the $\gamma^*p$ cross section into a universal
diffractive flux (depending only on $x_{I\!\!P}$) and a structure
function (depending only on $\beta$ and $Q^2$) is not tenable.  These
deviations from factorisation were demonstrated to be consistent with
an interpretation in which two individually factorisable components
contribute to $F_2^{D(3)}$.  These two components could be identified
with pomeron ($I\!\!P$) and meson contributions $\propto
\xpom^{-n_{I\!\!P}}$,$\xpom^{-n_{M}}$ where
$n_{I\!\!P}=1.29\pm0.03\pm 0.07$ and
$n_{M}=0.3\pm0.3\pm 0.6$.  Scaling violations, positive
with increasing ${\rm log}Q^2$ for all $\beta$, were observed and
could be interpreted in terms of a large gluon component in the
diffractive exchange, concentrated near $x_{g/I\!\!P}=1$ 
at $Q^2\sim 2.5\,{\rm GeV^2}$~\cite{QCD93,H1PARIS,H1F2D94,H1WARSAW}.

Given the significant progress in the understanding of diffractive
dynamics that has been achieved with the existing data, the goal of
future measurements is twofold: to extend the kinematic regime of
existing measurement, and to achieve highest possible precision,
particularly where existing measurements have uncovered 
interesting physics. In the 1994
measurement, with $5$ bins per decade in both $Q^2$ and $\xpom$, and
$7$ bins in $\beta$ between $0.01$ and $0.9$, then in the interval
$8.5<Q^2<65\,{\rm GeV^2}$ there were an average of $\sim100$ events
per bin, corresponding to a statistical accuracy of $\sim
10\%$\footnote{The variation of the cross section with $Q^2$ of
  approximately $Q^{-4}$ was partially offset by an increasing bin
  size with increasing $Q^2$.}. The different sources of systematic
error for the 1994 measurement are shown in table \ref{tab:syst},
along with an estimate of the level to which they may be reduced in
the future.
\begin{table}  
\begin{small}
\begin{center}
\begin{tabular}{|c|c|c|}\hline
\multicolumn{2}{|c|}{H1 1994 Preliminary} & 
H1 Future  \\ \hline
Error Source & $\delta F_2^{D(3)}/F_2^{D(3)}$ & 
$\delta F_2^{D(3)}/F_2^{D(3)}$ \\ \hline
Main Calo' Hadronic E Scale: $\pm 5\%$ & $3\%$  & 
 $\lapprox 1\%$ \\ \hline
Backward Calo' Hadronic E Scale: $\pm 15\%$ & $3\%$ & 
 $\lapprox 0.5\% $   \\ \hline
Backward Calo' Elec. E Scale: $\pm 1.5\% $ & $5\%$ & 
 $\lapprox 3\%$ \\ \hline
Tracking Momentum Scale $\pm3\%$ & $2\%$ & 
 $\lapprox 1\%$ \\ \hline
 Scattered Lepton Angle: $\pm 1\,{\rm mrad}$ & 
 $2\%$ & $1\%$ \\ \hline
 $t$ dependence: $e^{6\pm2t}$ & $1\%$ & 
 $0.5\%$ \\ \hline
 $x_{I\!\!P}$ dependence: $\xpom^{n\pm0.2}$ & $3\%$  & 
 $1\%$ \\ \hline
 $\beta$ dependence   & $3\% $ &
 $1\%$ \\ \hline
 $M_X$/$x_{I\!\!P}$ resolution & $4\%$ & $2\%$ \\ \hline
 Background (photoproduction and non-$ep$) & $0.1\%$ & $<0.1\%$ \\ \hline
 MC Statistics/Model Dependence & $14\%$ & $4\%$ \\ \hline
\end{tabular}
\end{center}
\end{small}
\caption{\label{tab:syst} Sources of systematic error for the H1 1994 
Preliminary measurement of $F_2^{D(3)}$. The results of calculations 
to estimate the extent to which these uncertainties may be reduced
are shown in the right hand column. These calculations take cognisance
of the new SPACAL backward calorimeter installed by H1~\cite{SPACAL},
and rely upon future measurements made using the LPS and forward
neutron calorimeter based upon a minimum luminosity 
of $10\,{\rm pb}^{-1}$.}
\end{table}
  The largest single error arises from the combination of limited Monte 
  Carlo statistics ($10\%$) and different possible final state topologies 
  leading to varying corrections for finite efficiency and 
  resolution ($10\%$). The latter contribution was estimated from the
  difference in the correction factors calculated for two 
  possible $\gamma^*I\!\!P$ interaction mechanisms:
\begin{itemize}
\item a quark parton model process in which the $\gamma^*$ couples
directly to a massless quark in the exchange with zero transverse momentum,
\item a boson--gluon fusion process in which the $\gamma^*$ couples
to a gluon in the diffractive exchange through a quark box.
\end{itemize}
  Numerous experimental measurements of diffractive final state 
  topologies at HERA are now available~\cite{H1WARSAW}
  which constrain the data to lie between these two possibilities.
  Therefore, it is reasonable to assume that the error arising from 
  a lack of knowledge of the final state topologies may be reduced by a 
  factor of $\sim2$ such that it no longer dominates the total error. 
  Monte Carlo statistics can obviously be increased. Therefore, it is 
  reasonable that in the future a measurement may be made with a total 
  systematic uncertainty of $5\%$ or less. To reduce the statistical
  error to $5\%$ ($3\%$) would require $8\,{\rm pb^{-1}}$ 
  ($22\,{\rm pb^{-1}}$) of data. 
 
  Less luminosity is required to achieve the same statistical precision 
  at lower $Q^2$. However, to reach the lowest possible $Q^2$ with 
  the widest possible range in $x$ (and hence in $\xpom$) it is 
  advantageous to run with the interaction vertex shifted forwards along
  the proton beam direction. We calculate that $2$~${\rm pb^{-1}}$ of such 
  data would give a  statistical accuracy of $5\%$ in the region of overlap
  between nominal and shifted vertex data, allowing a precise cross 
  check of the two analyses. It is important to note that
  the HERA magnets are not optimised for a shifted vertex configuration, 
  and so such data take twice as long to collect as for the nominal 
  configuration.   

  Theoretically, and consequently experimentally, the region of high 
  $\beta$ is of particular interest.
  The evolution of $F_2^{D(3)}$ with $Q^2$ at high $\beta$ is expected
  to depend crucially upon which evolution equations are pertinent to
  diffractive dynamics. In particular, a DGLAP~\cite{DGLAP} QCD 
  analysis~\cite{H1PARIS,H1F2D94,H1WARSAW} demonstrates that the H1 data are 
  consistent with a large gluon distribution, concentrated near
  $x_{g/I\!\!P}=1$ at $Q^2\sim2.5\,GeV^2$. In this case then $F_2^{D(3)}$
  should begin to fall with increasing $Q^2$ at $\beta=0.9$ for
  $Q^2\gg10\,{\rm GeV^2}$. The presence of a sufficiently large 
  ``direct'' term in the evolution equations would lead to a 
  indefinite increase with 
  $Q^2$~\cite{GS}. Thus a measurement 
  of $F_2^{D(3)}$ at high $\beta$ to the highest possible
  $Q^2$ is desirable.

  At fixed $ep$ centre of mass energy $\sqrt{s}$ the range of 
  $\beta$ that may be accessed decreases with increasing $Q^2$ such that:
\begin{equation}
\beta > \frac{Q^2}{s x_{I\!\!P}^{max}y^{max}}.
\end{equation}
  The acceptance for a measurement of  $F_2^{D(3)}$ at $\beta=0.9$, 
  based upon the interval $0.8<\beta<1$, therefore extends to 
  a maximum $Q^2$ of $\sim3500\,{GeV^2}$ for $x_{I\!\!P}<0.05$, where
  the cross section is likely to be dominantly diffractive. To 
  achieve $10\%$ statistical precision in a measurement in the interval
  $2000<Q^2<3500\,{\rm GeV^2}$ (bin centre $Q^2=2600\,{\rm GeV^2}$) 
  would require $\sim 200$~${\rm pb^{-1}}$. However, a measurement of 
  $30\%$ statistical precision, requiring only $\sim 20\,{\rm pb^{-1}}$, 
  would already be significant in light of theoretical calculations
  which differ by more than a factor of $2$ in this 
  region. 

\begin{boldmath}
\section{Measurement of $F_2^{D(4)}$}
\end{boldmath}
For particles with $E\ll E_P$ (here $E_P$ is the proton beam energy)
then the HERA magnets separate them from the proton beam allowing
particles to be tagged for $t\sim t_{min}$. For particles with $E\sim
E_P$ then only those particles with transverse momentum
$P_T^2\gapprox0.07\,{\rm GeV^2}$ may be tagged~\cite{ZEUS_LPS1}.
Consequently, for diffractive measurements then the acceptance in $t$
is limited to $t\lapprox -P_T^2/(1-\xpom)=-0.07\,{\rm GeV^2}$. For a
process with a highly peripheral $t$ dependence then this limitation
results in the loss of $30$ to $40\%$ of the total cross section.  The
geometrical acceptance of the current ZEUS detectors is $\sim
6\%$~\cite{ZEUSFPS}, giving an overall acceptance in the region
$\xpom<0.05$ of $\sim 4\%$.

 Tantalising first measurements of the $t$ dependence of deep--inelastic 
 events with a leading proton have been presented by the ZEUS 
 collaboration~\cite{ZEUS_LPS1}. The observed dependence on
 $t$ of ${\rm d}\sigma/{\rm d}t \propto e^{bt}$, 
 $b=5.9\pm1.2^{+1.1}_{-0.7}\,{\rm GeV}^{-2}$, lends strong 
 support to a diffractive interpretation of such interactions.
 The measurements of $b$ differential in $\xpom$, $\beta$ and $Q^2$ 
 that will be possible with increased luminosity are eagerly awaited.

 A preliminary measurement $F_2^{D(3)}$ using an LPS has also been
 made~\cite{ZEUS_LPS2}.  In order to use measurements of $F_2^{D(4)}$
 to estimate $F_2^{D(3)}=\int_{t_{min}}^{t_{max}}{\rm d}t\,F_2^{D(4)}$
 it is necessary to extrapolate from the measured region at high $|t|$
 into the region of no acceptance to $t_{min}$. To make this
 extrapolation with the minimal assumptions about the $t$ dependence,
 it is necessary to have at least three bins in $t$ for each bin in
 $\beta$, $Q^2$ and $\xpom$. This means that $\sim 150\,{\rm pb}^{-1}$
 of data are required to make a measurement of similar accuracy to the
 existing data for $F_2^{D(3)}$. Obviously it would be possible to
 study the $t$ dependence in aggregate volumes of phase space and make
 a measurements with a factor $3$ fewer statistics, relying upon model
 assumptions about the variation of the $t$ dependence with $\beta$,
 $Q^2$ and $\xpom$. Even with $150\,{\rm pb^{-1}}$ of data, the
 extrapolation into the $t$ region of no acceptance relies on the
 assumption that the $t$ dependence is the same in the measured and
 unmeasured regions. It is not clear that the resultant theoretical
 uncertainty will be less than the 5\% uncertainty in the overall
 normalisation resulting from proton dissociation for existing
 measurements of $F_2^{D(3)}$.

 The primary purpose of the LPS is the diffractive ``holy grail'' of
 measuring $F_2^{D(4)}(\beta,Q^2,\xpom,t)$. Measurements of any
 possible ``shrinkage'' of the forward diffractive scattering
 amplitude (increasing peripherality of the $t$ dependence with
 decreasing $\xpom$) are likely to have unique power for
 discriminating between different theoretical models of diffractive
 dynamics~\cite{MCDERMOTT}. In addition, the ability to select
 subsamples of events in which there is an additional hard scale at
 the proton vertex is of great theoretical interest~\cite{MCDERMOTT}.
 It is therefore of the utmost importance that the $100$ to $150\,{\rm
   pb^{-1}}$ of data necessary to make an accurate measurement of
 $F_2^{D(4)}(\beta,Q^2,\xpom,t)$ are collected by both collaborations
 with the LPS devices fully installed.

 The LPS also allows measurements at higher $\xpom$ for which the 
 rapidity gap between the photon and proton remnant systems $X$ and 
 $Y$ becomes too small to observe in the main calorimeter. 
 This will allow the contributions to the measured hadronic cross 
 section from subleading trajectories to be further investigated. 
 Combining information from the scattered lepton and proton will allow the 
 invariant mass $M_X$ to be reconstructed in the region $\xpom\gg0.05$, 
 where it would otherwise be impossible. 

 Tagged particles with $E\ll E_P^{beam}$ will provide information
 about the way in which protons dissociate into higher mass systems. 
 Particles thus produced are kinematically unlikely to generate 
 a significant background in the region $\xpom<0.05$, but for 
 larger $\xpom$ then this background will become increasingly important.
  

{\boldmath
\section{On the Determination of $F_L^{\pom}$}}
 A fundamental question in the study of hard diffractive processes is
 the extent to which perturbative QCD dynamics may be factorised from
 the proton vertex. Some recent calculations of diffractive cross
 sections~\cite{BUCHMUELLER,SCI} consider the interaction in terms of a
 hard (perturbative) phase producing a coloured partonic system which
 subsequently interacts with the colour field of the proton via the
 exchange of non-perturbative gluons. Such ``soft colour
 interactions''~\cite{SCI} allow colour to be exchanged between 
 the photon and proton remnant systems such that in some fraction 
 of the events both will become colour singlet states separated by 
 a potentially large rapidity gap. In such models the evolution of 
 the effective parton distribution functions (PDFs) will be driven by the 
 evolution of the PDFs of the proton. Alternatively, it is possible that 
 the presence of a large rapidity gap will confine any evolution dynamics
 to within the photon remnant system $X$. The effective PDFs will then 
 depend only upon $\beta$ and $Q^2$, and not upon $\xpom$ as in the 
 former case. 

\begin{figure}[htb]\unitlength 1mm
 \begin{center}
\begin{picture}(120,60)
   \put(-15,-75){\epsfig{file=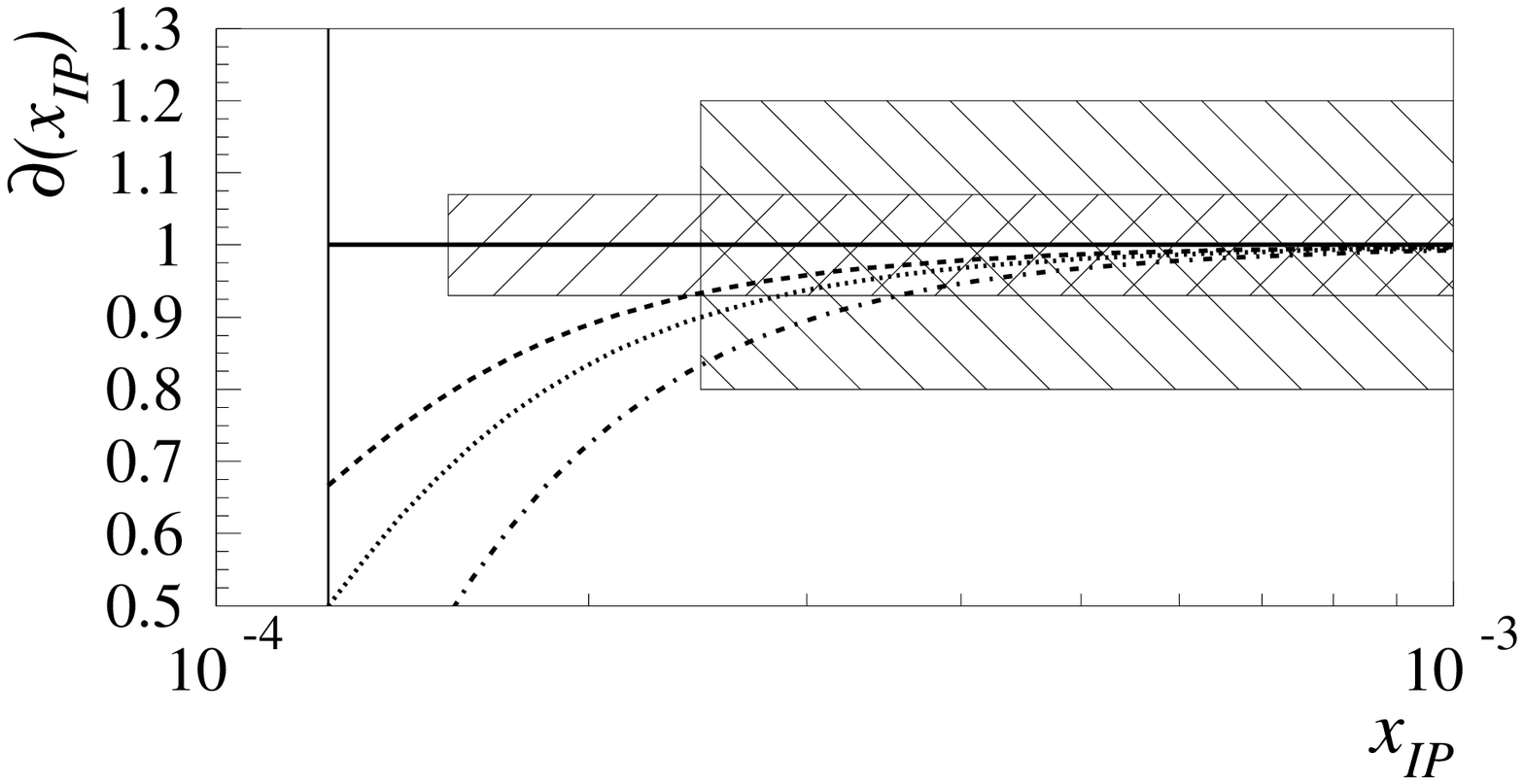,width=0.9\textwidth}}
{\footnotesize
\put(23,39){$0$}
\put(28,25){$0.5$}
\put(32,20){$1$}
\put(36,15){$R^{I\!\!P}=5$}
\put(20,58){$y=1$}}
\end{picture}
 \end{center}

 \caption{\label{fig:rpom} The expected fractional deviation $\delta(\xpom)$ 
of $F_2^{D(3)}$
from an extrapolation from high $\xpom$ for $R^{I\!\!P}=0$ (horizontal
solid line), $R^{I\!\!P}=0.5$ (dashed line), 
$R^{I\!\!P}=1.0$ (dotted line) and $R^{I\!\!P}=5.0$ (dash-dotted line)
for $Q^2=10\,{\rm GeV^2}$, $\beta=0.5$. The vertical solid line 
indicates the kinematic limit of $y=1$. The larger box 
indicates the area covered by the preliminary 1994 H1 
measurements, which extended
up to $y\sim 0.5$ with an accuracy of $\sim20\%$. The smaller box 
represents a future measurement with a total error of 7\% 
extending to $y=0.8$.}
\end{figure}

The 1994 H1 data have been demonstrated to be consistent with a factorisable 
approach in which a large gluon distribution is attributed to the 
pomeron in the region $\xpom<0.01$ or over the whole kinematic 
range for the sum of two individually factorisable 
components~\cite{H1WARSAW}. At next to 
leading order (NLO) then a large gluon distribution 
$G^{I\!\!P}(\beta,Q^2)$ necessitates
a large longitudinal structure function, $F_L^{I\!\!P}(\beta,Q^2)$:
\begin{equation}
F_L^{I\!\!P}(\beta,Q^2) = \frac{\alpha_s(Q^2)}{4\pi}\cdot 
\beta^2\int_{\beta}^{1}\frac{{\rm d}\xi}{\xi^3}
\left[\frac{16}{3}F_2^{I\!\!P}(\xi,Q^2) + 8\sum_{i}e_i^2(1-\frac{\beta}{\xi})
\,\xi G^{I\!\!P}(\xi,Q^2)\right].
\end{equation}
A prediction for $F_L(\beta,Q^2)$ based upon a NLO QCD analysis of the 
data could be tested directly since the wide range in $\xpom$ accessible 
at HERA leads to a wide variation in the $eI\!\!P$ 
centre of mass energy $\sqrt{s_{eI\!\!P}}=\sqrt{\xpom s}$. This means that
in factorisable models, at fixed $\beta$ and fixed $Q^2$ the same 
partonic structure of the pomeron may be probed at different values of 
 $\xpom$, corresponding to different values of $y$:
\begin{equation}
 y = \frac{Q^2}{s\beta} \cdot \frac{1}{\xpom}.
\end{equation}
If the dependence of the diffractive structure function can be factorised
such that 
\begin{eqnarray}
F_2^{D(3)}(\beta,Q^2,\xpom)=f(\xpom) F_2^{I\!\!P}(\beta,Q^2)
\end{eqnarray}
and $F_L^{I\!\!P}$ is non-zero, then a measurement of
$F_2^{D(3)}$ assuming $F_L^{I\!\!P}$=0 is lower than the correct value
of $F_2^{D(3)}$ by the factor $\delta(y,\beta,Q^2)$:
\begin{equation}
\delta = \frac{F_2^{D(3)}(Measured)}{F_2^{D(3)}(True)}
= \frac{2(1-y) + y^2/[1+R^{I\!\!P}(\beta,Q^2)]}{2(1-y)+y^2}
\end{equation}
 Thus under the assumption of factorisation,  a measurement of 
 $F_L^{I\!\!P}(\beta,Q^2)$ is possible by measuring the extent of the 
 deviation from a simple $\xpom^{-n}$ extrapolation from higher $\xpom$. 
 The size of this effect is shown for different values of 
 $R^{I\!\!P}=F_L^{I\!\!P}/(F_2^{I\!\!P}-F_L^{I\!\!P})$ 
 in figure \ref{fig:rpom}. 
 Also shown is the range covered by the existing 1994
 data and the range that could be covered by future measurements at HERA. 
 It is clear that in order to see this effect it is necessary to
 make a measurement at $y$ values above $0.7$ with a total error of less
 than $10\%$. For high values of $\beta$ the extrapolation 
 of $F_2^{D(3)}$ into the high $y$ region of interest should not be 
 significantly compromised by the possible presence of subleading 
 trajectories, which contribute only at high $\xpom$ (low $y$). 

 A comparison between the values of  $F_L^{I\!\!P}$ determined from 
 such apparent deviations from factorisation and the values 
 expected from a QCD analysis of $F_2^{D(3)}$ constitutes a powerful
 test of both the validity of factorisation and the applicability of
 NLO QCD to diffraction at high $Q^2$. 

{\boldmath
\section{Measurements of $R^{D(3)}$ and $R^{D(4)}$}}
 Since an evaluation of $F_L^{I\!\!P}$ using the techniques 
 described in the previous section require  theoretical 
 assumptions concerning factorisation, such an analysis is 
 clearly no substitute for a direct measurement of the ratio of
the longitudinal to transverse diffractive cross sections, $\Rd(\xpom,
\beta)$. A good measurement of this quantity is vital for a full
understanding of the diffractive mechanism and should provide an
exciting testing ground for QCD. There is at present no theoretical
consensus on what values to expect for $R^{D}$, although all models
suggest a substantial dependence on $\beta$ with most suggesting an
extreme rise as $\beta \rightarrow 1$~\cite{MCDERMOTT}. 
A measurement of $R^{D}$ to any precision leads us  into unexplored 
territory.

Measurements of $R^D$ have so far been restricted to DIS exclusive
vector mesons production~\cite{VDM} by a direct measurement of the
polarisation of the final state resonance.  This method could perhaps
be used for the bulk data if the directions of the final state partons
could be inferred, but is likely to be difficult due to the problems
of running jet analyses on low mass final states.  Instead we
investigate a slightly modified version of the method used to
determine $R(x, Q^2)$ for inclusive DIS~\cite{FLHWS}.

The general form relating the structure functions $F^D_2$ and $F^D_1$
to the $ep$ differential diffractive cross section can be written in
analogy to the inclusive cross sections~\cite{GUNNAR}.
\begin{eqnarray}
  \frac{{\rm d}^4 \sigma^D_{ep}}{{\rm d} x_{I\!\!P}\,{\rm d} t\,{\rm
      d} x\,{\rm d}Q^2}=
  \frac{4\pi\alpha^2}{xQ^4}(1-y+\frac{y^2}{2(1+R^{D(4)}(x,Q^2,x_{I\!\!P},t))}){F_2}^{D(4)}(x,Q^2,x_{I\!\!P},t),
\label{eqn:sig4}
\end{eqnarray}
where
$R^{D(4)}=(F_2^{D(4)}-2xF_1^{D(4)})/(2xF_1^{D(4)})=\sigma^D_L/\sigma^D_T$.
Although a measurement of $R^{D(4)}$ as a function of all 4 variables
is the most desirable measurement and must be an experimental goal,
statistical limitations are likely to mean that initial measurements
must be made without a reliance on a leading proton spectrometer (LPS) and so
$t$ is not measured. In this case we define $F_2^{D(3)}$ and
$R^{D(3)}$ as
\begin{eqnarray}
\frac{{\rm d}^3 \sigma^D_{ep}}{{\rm d} x_{I\!\!P}\,{\rm d} x\,{\rm d}Q^2}= 
\frac{4\pi\alpha^2}{xQ^4}(1-y+\frac{y^2}{2(1+R^{D(3)}(x,Q^2,x_{I\!\!P}))}){F_2}^{D(3)}(x,Q^2,x_{I\!\!P}),
\label{eqn:sig3}
\end{eqnarray}
In this case $R^{D(3)}$ is the ratio of the longitudinal to transverse
cross section only if $R^{D(4)}$ has no dependence on $t$. 

Analysis of equation~\ref{eqn:sig3} reveals that in order to make a
measurement of $R^{D}$ independent of $F^D_2$ at least two $ep$ cross
sections must be compared at the same values of $x$, $Q^2$ and $\xpom$
but different values of $y$. This is achieved by varying the $ep$
centre of mass energy, $\sqrt{s}$.  There are of course many possible
running scenarios for which either or both beam energies are changed
to a variety of possible values. A full discussion on this point is
given in~\cite{FLHWS}.  For the present study we examine the case when
the proton beam is roughly halved in energy from 820~GeV to 500~GeV
and the electron beam remains at a constant energy of $27.5$~GeV so
that data is taken at the 2 values of $s$ of $s=90200$~${\rm GeV}^2$
and $s=55000$~${\rm GeV}^2$.  This setup allows for a reasonable
luminosity at the low proton beam energy and enables systematic
uncertainties concerned with detection of the scattered electron to be
minimised. In this scheme we make a measurement of the ratio of the
$ep$ differential cross sections, $r=\sigma^D_{hi}/ \sigma^D_{lo}$, for
two values of $y$, $y_{hi}$ and $y_{lo}$ (corresponding to the high
and low values of $s$) for fixed $x$, $Q^2$, $\xpom$ and (if measuring
$R^{D(4)}$) $t$. Equation~\ref{eqn:sig4} or \ref{eqn:sig3} is then
used to determine $\Rd$.

\begin{figure}[htb]
 \begin{center}
 \epsfig{file=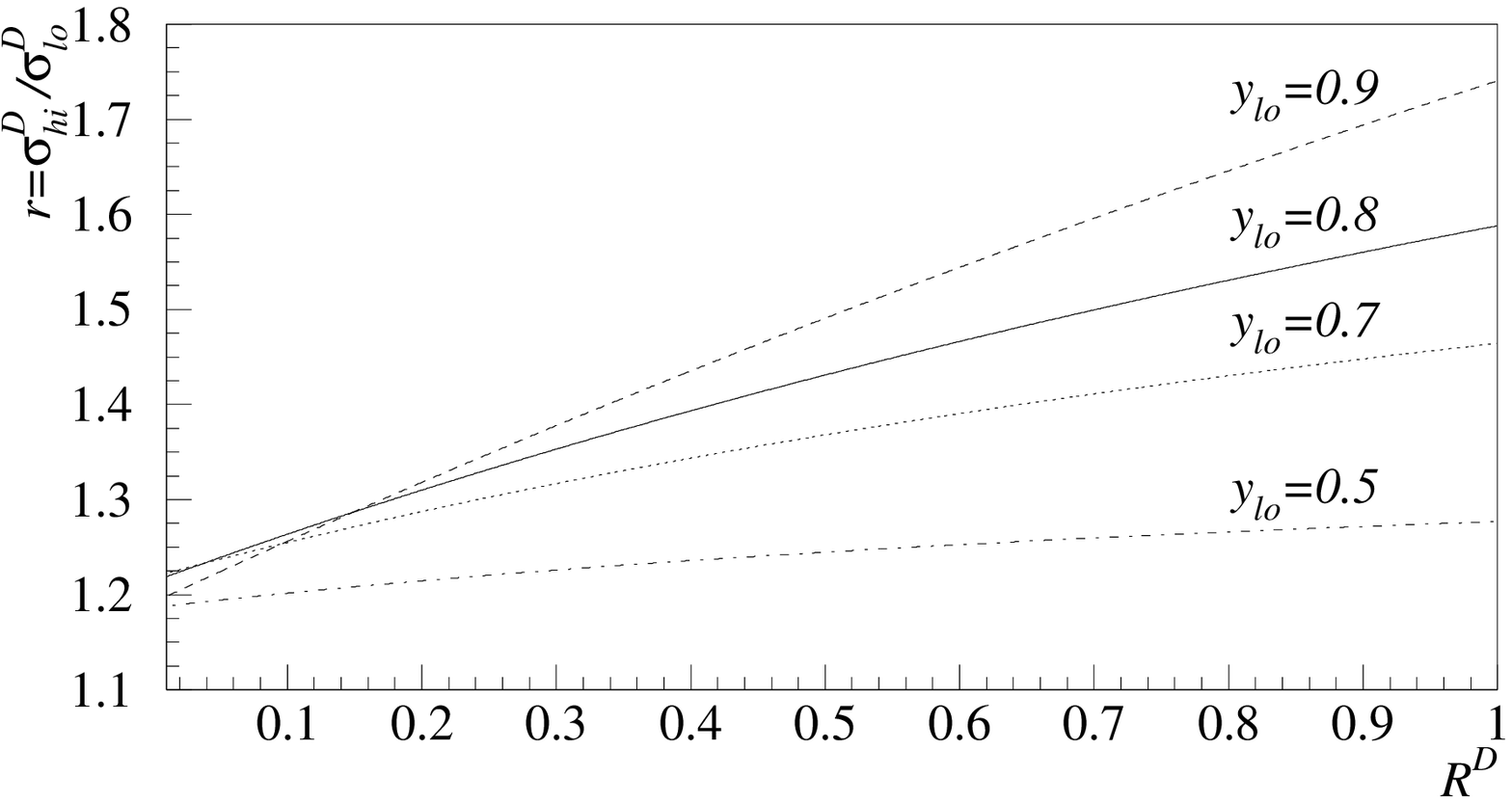,width=0.7\textwidth}
 \end{center}
  \caption  {Dependence of the ratio, $r$,  of the $ep$ cross sections at
$s=90200$ and $s=55000$ with $R^D$ for various values of $y$ at $s=55000$.} 
  \label{fig:flycut}
\end{figure}

It is also apparent from equation~\ref{eqn:sig3} that in order to have
the greatest sensitivity to $R^{D}$ measurements must be made at the
highest $y_{lo}$ possible (and thus lowest electron energy). This is
illustrated in figure~\ref{fig:flycut}, where it can be seen that for
values of $y_{lo}=0.5$ (or lower) there is little change of $r$ for
different values of $R^{D}$.  The upper limit in $y_{lo}$ is crucially
dependent on the ability of the detectors to discriminate and veto
against photoproduction events in which a pion is misidentified as an
electron. Experience has shown, however, that for the diffractive
events the low mass of the final state reduces the chance of faking
electron when compared to the more energetic non-diffractive events.
For this study we take a central value of $y_{lo}=0.8$ with a lower
(upper) bin limit of 0.75 (0.85) so that good electron identification
for energies above 4.15~GeV is assumed.

\begin{figure}[htb]
 \begin{center}
 \epsfig{file=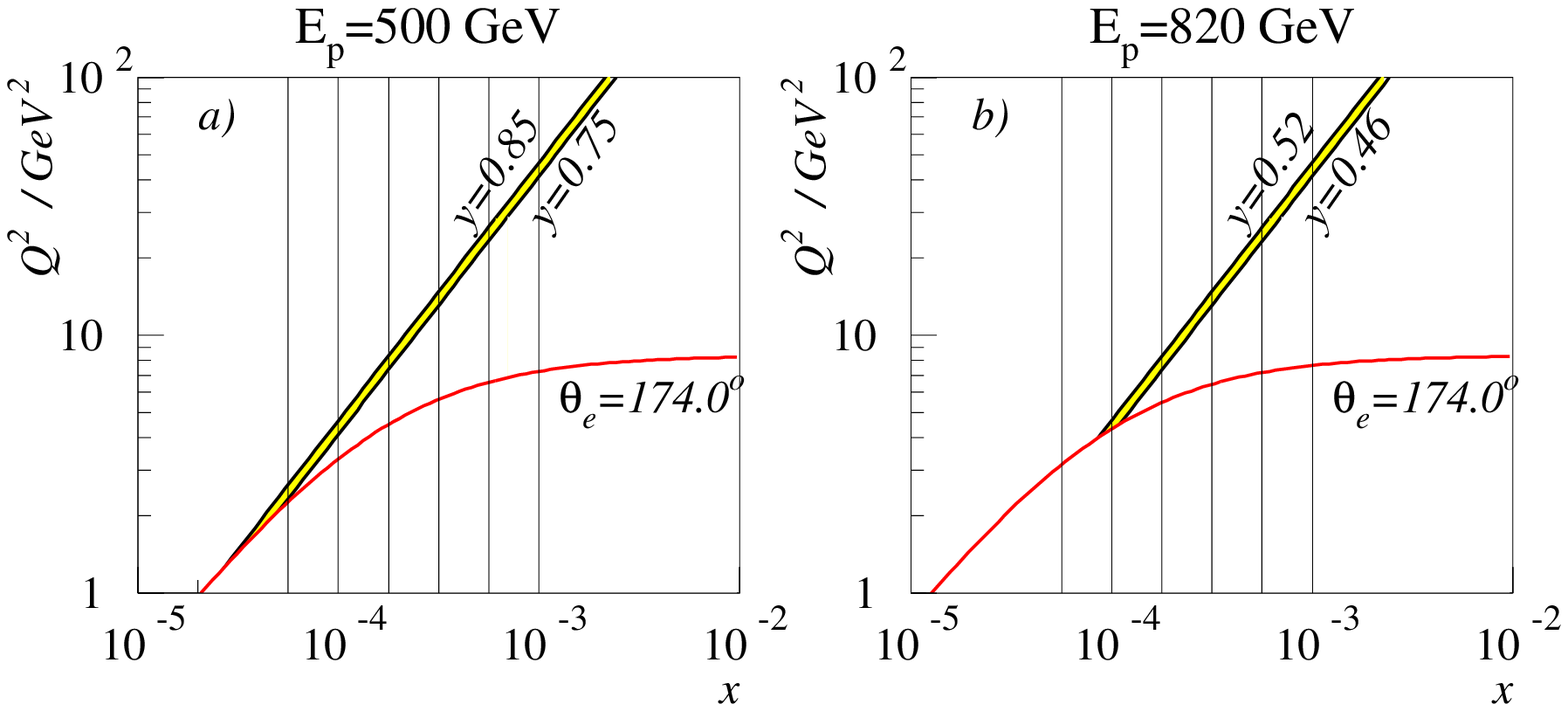,width=\textwidth}
 \end{center}
 \vspace{-1cm}
  \caption  {A projection of the kinematic range onto the $x$--$Q^2$ plane
    for a) a proton beam of $E_p=500$~GeV and b) a proton beam of
    $E_p=820$~GeV for an electron beam of 27.5~GeV. The shaded region
    represents the proposed region of study. Also shown is the
    restriction on the kinematic range imposed by a maximum $\theta_e$
    cut of $174^\circ$.}
  \label{fig:flbins}
\end{figure}

The kinematic range of the measurement projected onto the $x$--$Q^2$
plane is shown in figure~\ref{fig:flbins} for both CMS energies.  To
ensure that the scattered electrons are well contained within the
backward detectors we place a maximum $\theta_e$ cut of $174^\circ$.
This restricts us to $Q^2>5$~${\rm GeV}^2$ and $x>10^{-4}$.  In order to
ensure good acceptance in the forward region we impose a cut
of $\xpom<0.01$.

For low electron energies the kinematic variables are well
reconstructed and have good resolutions if the electron only method
is used~\cite{H1F2,ZEUSF2}.  Since the major problem with this
measurement will be in reducing the statistical error we select bins
as large as possible whilst maintaining enough bins to investigate any
variation of $R^D$ with the kinematic quantities. A suitable choice
would be 4 bins per decade in $x$, 4 bins in $\beta$ and if the LPS
is used 2 bins in $t$. The bins in $\beta$ and $t$ are optimised so as
to contain approximately equal numbers of events at each $x$ value.
Identical bins in these variables are used for both CMS energies.

In order to estimate the statistical errors on the measurement we used
the RAPGAP generator~\cite{RAPGAP} with a fit to the measured H1
$F_2^{D(3)}$~\cite{H1F2D94} at $s=90200$~${\rm GeV}^2$ and used
equation~\ref{eqn:sig4} to determine the statistics at $s=55000$~${\rm
  GeV}^2$. We assumed 100\% efficiency for measurements made without
any LPS and 4\% efficiency for those with. The expected number of
events per integrated luminosity in each bin is summarised in
table~\ref{tab:noevents} for an example $R^D=0.5$.

\begin{table}
\begin{center}
\begin{tabular}{|r|c|c|c|c|} \hline
 & \multicolumn{2}{c|}{Number of Events} &
 \multicolumn{2}{c|}{Number of Events} \\ 
 $\log_{10} x$ & \multicolumn{2}{c|}{without LPS} &
 \multicolumn{2}{c|}{with LPS} \\  \cline{2-5}
   & $E_P=820$~GeV & $E_P=500$~GeV & $E_P=820$~GeV & $E_P=500$~GeV \\ \hline
-4.125  & 41 & 28 & 0.82 & 0.56 \\ \hline
-3.875  & 36 & 25 & 0.72 & 0.50\\ \hline
-3.625  & 19 & 13 & 0.38 & 0.27\\ \hline
-3.375  & 9  & 6  & 0.18 & 0.12\\ \hline
\end{tabular}
\end{center}
\caption{The estimated number of events in each bin 
  for an integrated luminosity of 1~${\rm pb}^{-1}$ for the 2 proton
  beam energies and an electron beam energy of $27.5$~GeV, assuming 4
  bins per decade in $x$, 4 bins in $\beta$ and (for measurements with
  the LPS) 2 bins in $t$. $R^D$ was set to 0.5.}
\label{tab:noevents}
\end{table}

For systematic errors we estimate an error of $\delta(r)/r$ of 5\%.
This error is conservatively evaluated by taking the estimated error
on $F^D_2$ (see above) and assuming any improvement that arises from
taking a ratio is offset by increased uncertainty in the
photoproduction background and radiative corrections.

\begin{figure}[htb]
 \begin{center}
 \epsfig{file=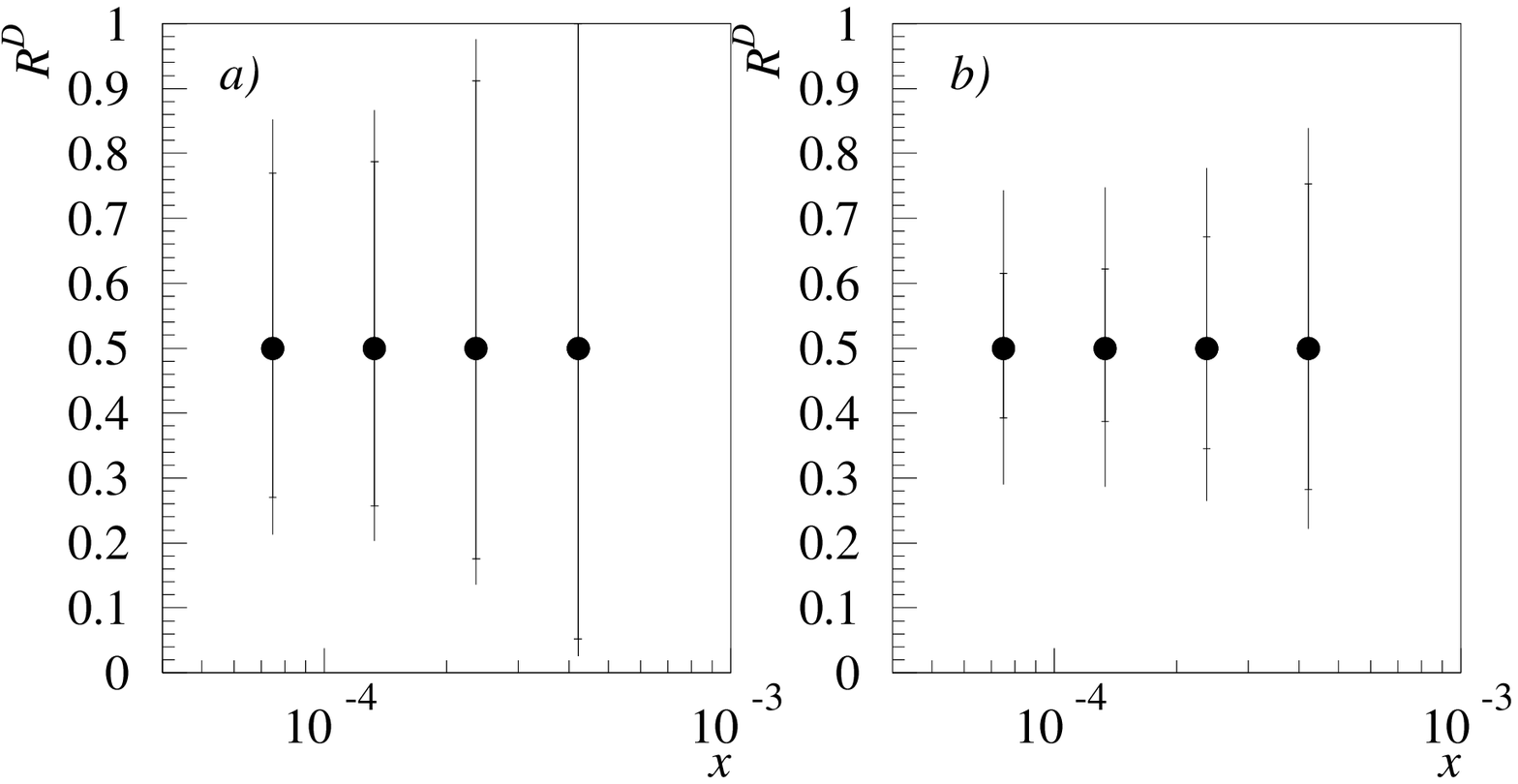,width=0.7\textwidth}
 \end{center}
  \caption{The estimated errors for an example central value
    of $R^D=0.5$ for a) 10(500)~${\rm pb}^{-1}$ at $s=55000$~${\rm
      GeV}^2$ and 50(2500)~${\rm pb}^{-1}$ at $s=90200$~${\rm GeV}^2$
    and b) 50(2500)~${\rm pb}^{-1}$ at $s=55000$~${\rm GeV}^2$ and
    250(12500)~${\rm pb}^{-1}$ at $s=90200$~${\rm GeV}^2$ for a
    measurement of $R^{D(3)}$ ($R^{D(4)}$). The inner error bar
    represents the statistical error and the outer the statistical and
    systematic error added in quadrature.}
  \label{fig:flrlum}
\end{figure}

An example of the sort of precision which may be obtained for a
measurement of $R^{D(3)}$ and $R^{D(4)}$ is shown in
figure~\ref{fig:flrlum}. For this study we assumed that many more data
would be obtained at the high CMS energy.  It can be seen that for an
integrated luminosity of 10~${\rm pb}^{-1}$ at the lower $s$ value a
measurement of $R^{D(3)}$ is statistically dominated with an error 
around 60\% if $R^{D(3)}=0.5$ for the lowest value of $x$. For an integrated
luminosity of 50~${\rm pb}^{-1}$ at the lower $s$ value statistical
and systematic errors become comparable and $R^{D(3)}$ can be measured
to 40\% accuracy. For measurements of $R^{D(4)}$ very high integrated
luminosities are required -- at least a factor of 50 is needed for a
similar precision to $R^{D(3)}$.

\boldmath
\section{Measuring $F_{2~{\it charm}}^{D}$}
\unboldmath

Since the leading mechanism in QCD for the production of charm quarks
is the boson gluon fusion process, the diffractive charm structure 
function $F_{2~{\it charm}}^{D}$ is very sensitive to the gluonic component 
of the diffractive exchange. It is important to establish whether 
the measured $F_{2~{\it charm}}^{D}$ is consistent with that 
expected from a QCD analysis of the scaling violations in $F_2^{D(3)}$.
In addition, it has already been observed in the photoproduction
of $J/\psi$ mesons that the charm quark mass provides a sufficiently 
large scale to generate the onset of hard QCD dynamics. The extend to 
which the charm component of $F_2^{D(3)}$ exhibits a different 
energy ($\xpom$) dependence to that of the total will provide insight 
into the fundamental dynamics of diffraction.

The method used here for tagging charm events uses the $D^{*+}$
decay\footnote{Charge conjugate states are henceforth implicitly included.}
$D^{*+} \rightarrow D^0 \slowpi^+ \rightarrow (K^-\pi^+) \slowpi^+$.
The tight kinematic constraint imposed by the small difference between
the $D^{*+}$ and $D^0$ masses means that the mass difference
$\Delta M = M(K\pi\slowpi) - M(K\pi)$ is better resolved than the
individual masses, and the narrow peak in $\Delta M$ provides a clear
signature for $D^{*+}$ production.  The chosen $D^0$ decay mode is the
easiest to use because it involves only charged tracks and because the
low multiplicity means that the combinatorial background is small
and that the inefficiency of the tracker does not cause a major problem.

\begin{figure}[htb]
 \setlength{\unitlength}{1cm}
 \begin{picture}(16.0,5.5)
  \put(0.0,0.0){\epsfig{file=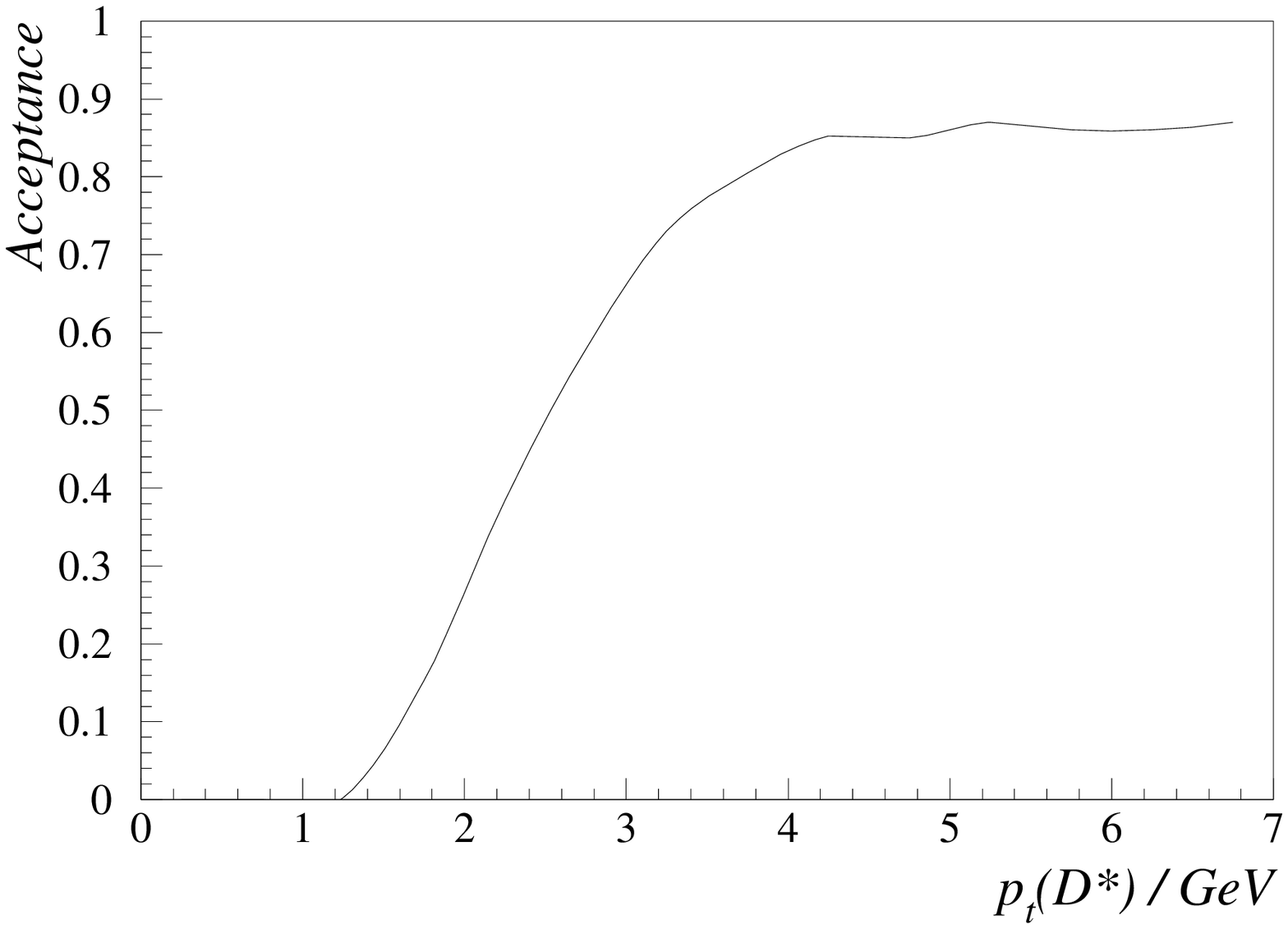,height=5.5cm}}
  \put(8.0,0.0){\epsfig{file=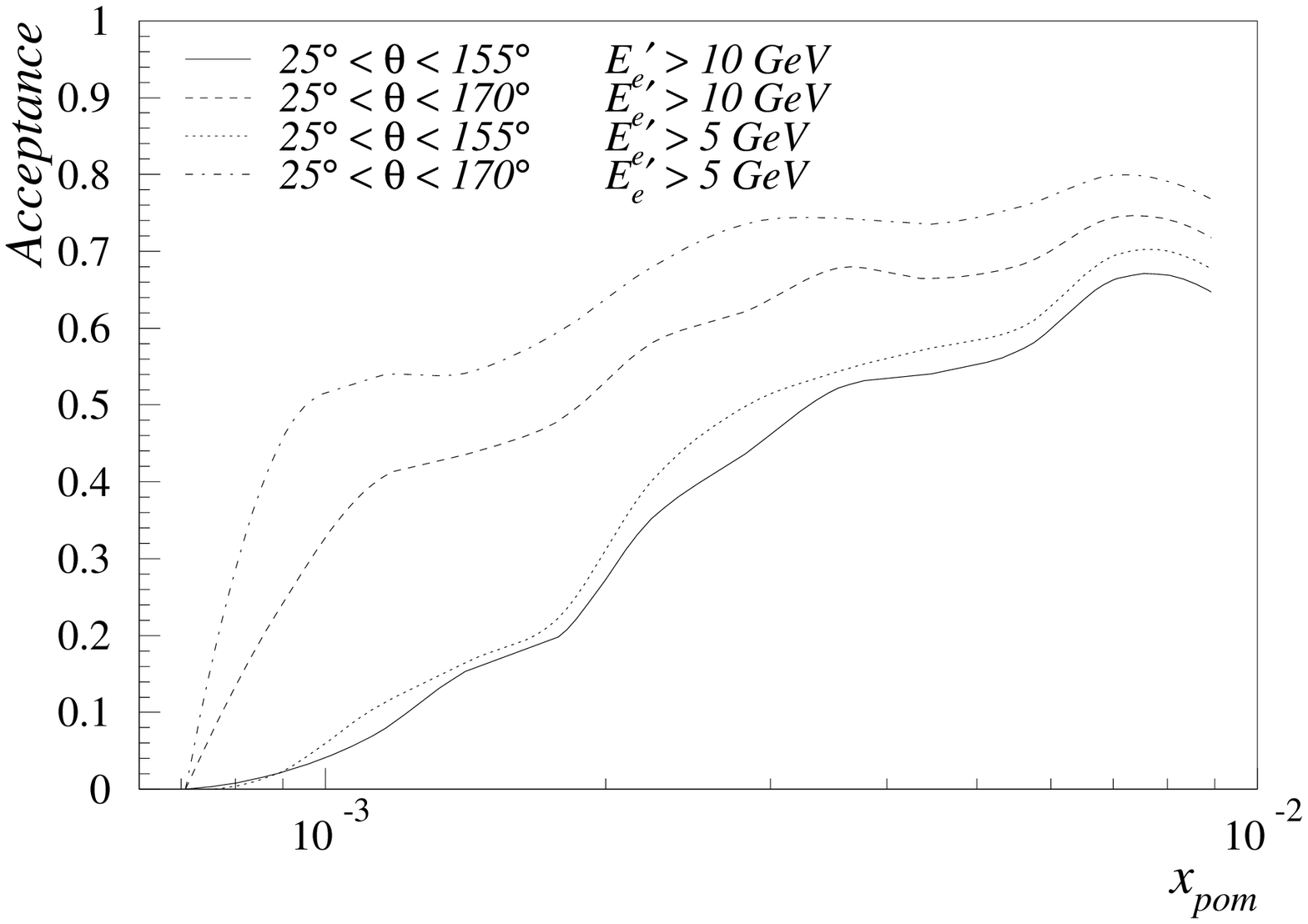,height=5.5cm}}
  \put(6.9,5.0){\small (a)}
  \put(14.9,5.0){\small (b)}
 \end{picture}
 \caption{\label{fig:acc}
          The acceptance for $D^{*+}$ in the region
          $10~\gevsq<Q^2<100~\gevsq$, shown as a function of
          (a)~$p_{\perp}$ and (b)~$\xpom$.
          The continuous line in (b) shows the results with central tracking
          only and a requirement $E_e^{\prime}>10$~GeV.
          The other lines show the effect of extending track
          coverage in the backward direction and including $E_e^{\prime}$
          down to 5~GeV.
         }
\end{figure}

A prediction of the observed number of events is obtained using
RAPGAP with a hard gluon dominated
pomeron structure function taken from a QCD analysis of
$F_2^{D(3)}$~\cite{H1F2D94}. 
 The cross section predicted by this model for $D^{*\pm}$ production
 in diffractive DIS is compatible with the value measured
 in~\cite{H1WARSAW}.
The acceptance of the detector is simulated by applying cuts on
the generated direction ($\theta$) and transverse momentum ($p_{\perp}$)
of the decay products and on the energy of the scattered lepton
($E_e^{\prime}$).
The $p_{\perp}$ cut used is 150~MeV, which is approximately the
value at which the H1 central and forward trackers reach full efficiency.
This cut has a major influence on the acceptance for $D^{*+}$ mesons,
because the momentum of the slow pion $\slowpi$ is strongly correlated
with that of the $D^{*+}$, so a $D^{*+}$ with $p_{\perp}$ much less than
150~MeV$ \times M_{D^{*+}} / M_{\pi^+} \approx 2$~GeV cannot be
detected.  The $p_{\perp}$-dependence of the acceptance is shown in
figure~\ref{fig:acc}a.
There is no obvious way of extending the tracker acceptance to lower
$p_{\perp}$, so this cut is not varied.

\begin{figure}[htb]
 \setlength{\unitlength}{1cm}
 \begin{picture}(16.0,5.5)
  \put(0.0,0.0){\epsfig{file=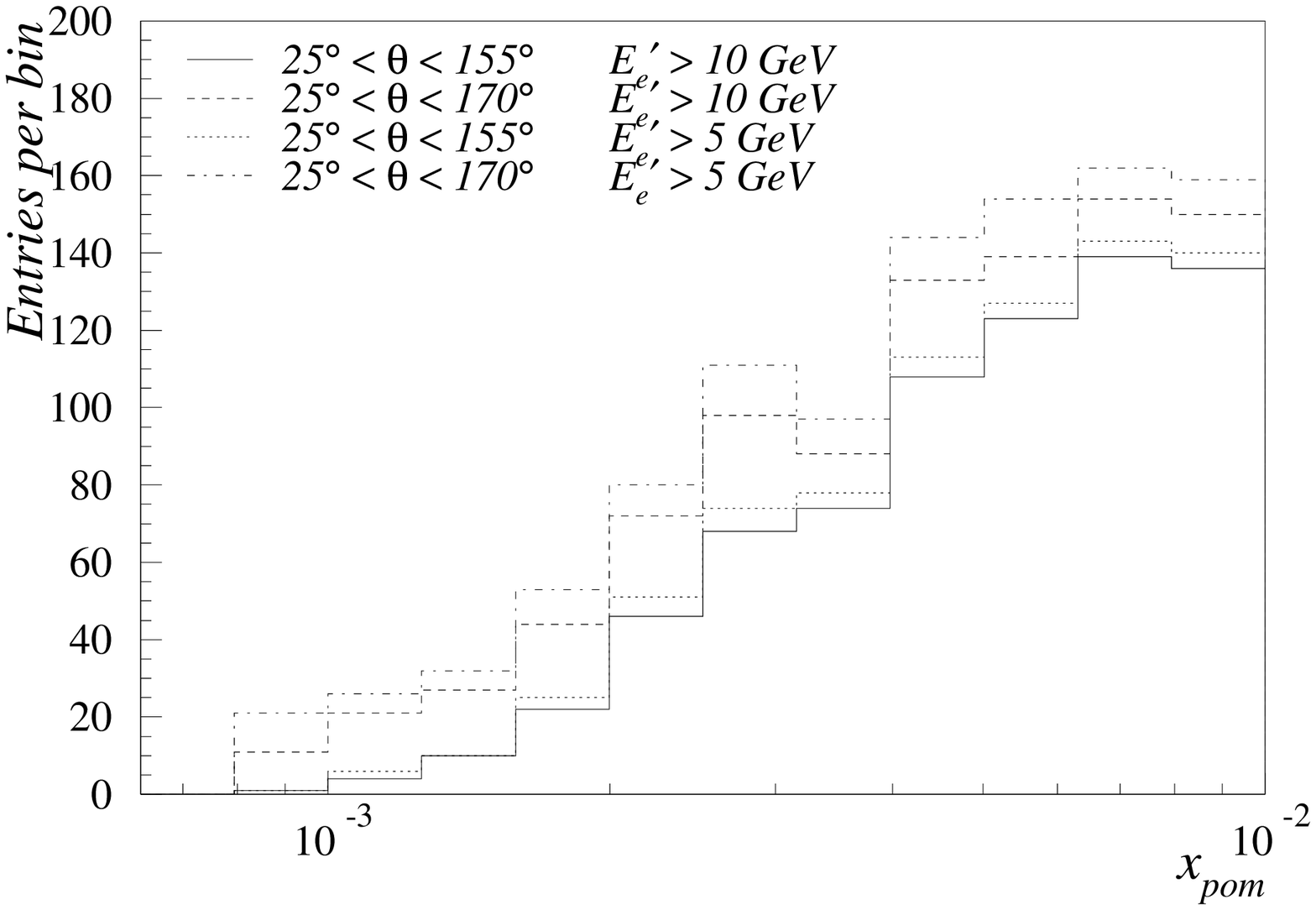,height=5.5cm}}
  \put(8.0,0.0){\epsfig{file=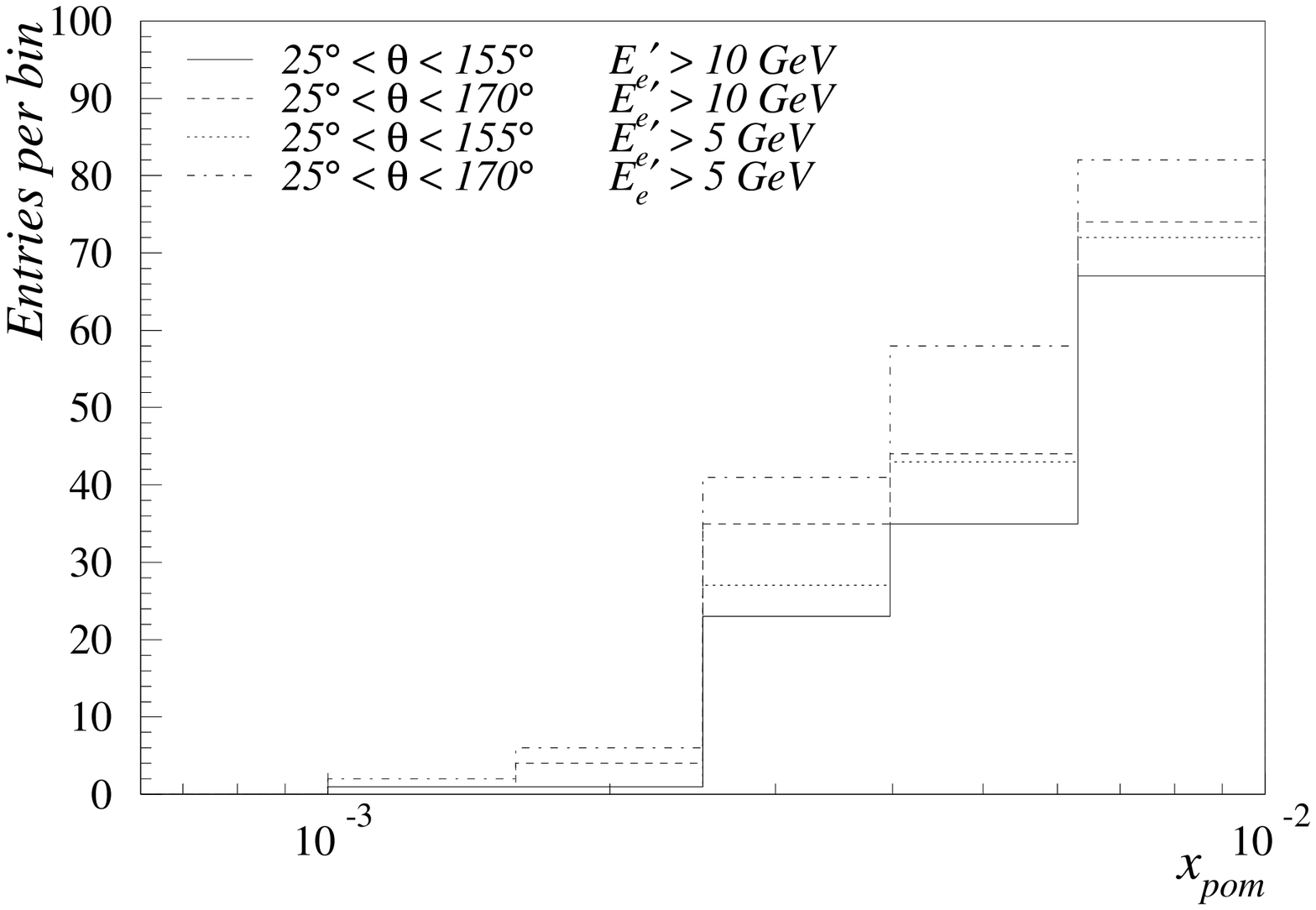,height=5.5cm}}
  \put(6.9,5.0){\small (a)}
  \put(14.9,5.0){\small (b)}
 \end{picture}
 \caption{\label{fig:num}
          The number of $D^{*\pm}$ expected to be observed in
          750~pb$^{-1}$ in the range
          (a)~$10~\gevsq<Q^2<25~\gevsq$ and
          (b)~$50~\gevsq<Q^2<100~\gevsq$,
          predicted using RAPGAP with a hard-gluon-dominated
          pomeron.}
\end{figure}

Figure~\ref{fig:acc}b shows the average acceptance for a $D^{*+}$ over
the region $10~\gevsq<Q^2<100~\gevsq$ and all values of $\beta$
and for $p_{\perp}>2\,{\rm GeV}$. It can be seen that extending the 
angular coverage from the present $25<\theta<155^{\circ}$ range 
in the backward direction to $170^{\circ}$ in conjunction with 
lowering the scattered lepton energy cut used in present analyses 
significantly improves the acceptance, especially at low $\xpom$.
Figure~\ref{fig:num} shows the number
of $D^{*\pm}$ which one might expect to detect in the low- and high-$Q^2$
regions with a total integrated luminosity of 750~pb$^{-1}$. 
It can be seen that even with this large integrated luminosity, 
cross section measurements can only be made with an accuracy of 10\%
in this binning.

Whilst one waits for $750\,{\rm pb^{-1}}$, it may be worthwhile
attempting to increase the statistics by investigating other decay 
modes which are experimentally more demanding. 
The $D^{*+}$ decay to $D^0 \slowpi^+$ has a branching fraction of
nearly 70\%~\cite{pdg96} and is the only decay mode giving
a charged track in addition to the $D^0$ decay products.
However, the $D^0$ decay to $K^-\pi^+$ has a branching fraction
of slightly less than 4\%~\cite{pdg96}, so there is
clearly room for a large improvement in statistics if other channels
can be used.  
For example the use of a silicon vertex detector close to the 
interaction point,
such as that already partly installed in H1, should enable the secondary
vertex from the decay of the $D^0$, which has a decay length
$c\tau = 124~\mu$m~\cite{pdg96}, to be tagged.  This could be used to
extend the analysis to other channels, including semileptonic decays,
which are otherwise difficult to use. 
The gain in statistics, neglecting
the vertex-tagging inefficiency, can be up to a factor of $\sim 20$
(if all channels are used), with a further factor of $\sim 2$ available
if inclusive $D^0$ production is used rather than relying on the $D^{*+}$
decay.

\section{Summary and Conclusions}
  We have argued that a precise hadron level definition of the cross section 
  to be measured is essential in order that future high precision
  measurements of diffractive structure functions may be treated in
  a consistent theoretical manner. Although $20\,{\rm pb}^{-1}$ of
  integrated luminosity will be enough to achieve precision in 
  the measurement of $F_2^{D(3)}$ at moderate $Q^2$, in excess 
  of $100\,{\rm pb^{-1}}$ is necessary to make a comprehensive
  survey of diffractive structure within the kinematic limits of HERA.
  An attempt to determine $F_L^{I\!\!P}$ will be an important
  tool in establishing the validity of both factorisation and 
  NLO QCD in diffractive interactions. A direct measurement of 
  $R^{D(3)}$ is shown to be feasible with $10\,{\rm pb^{-1}}$ of 
  integrated luminosity taken at lower proton beam energy. 
  A substantial integrated luminosity is demonstrated to 
  be necessary to complete an exhaustive study of the diffractive 
  production of open charm, although statistics can be markedly
  improved by exploiting the full range of possible charm decays.



\begin{thebibliography}{99}

\bibitem{H1F2D93} H1 Collaboration, T.~Ahmed et ~al., 
Phys.~Lett.~{\bf B348}(1995) 681-696. \\
ZEUS Collaboration, M.~Derrick et al.~, Z.~f.~Physik~{\bf C68}(1995). 

\bibitem{ZEUS_MASS93}
ZEUS Collaboration, M.~Derrick et al.~, Z.~f.~Physik~{\bf C70}(1996). 

\bibitem{H1F2D94}
A.~Mehta, Proceedings of the Topical Conference on Hard Diffractive 
Processes, Eilat, Israel, February 1996. \\
P.~Newman, Proceedings of the International Workshop on Deep--Inelastic 
Scattering and Related Phenomena, Rome, Italy, April 1996, DESY-96-162. 

\bibitem{H1WARSAW}
J.~Phillips, Proceedings of the {\em XXVIII} International Conference
on High Energy Physics, Warsaw, Poland, July 1996. See also contributed
papers {\bf pa02-061}, {\bf pa02-060}, {\bf pa02-063} and {\bf pa02-068}.

\bibitem{MCDERMOTT} 
  M.~McDermott, G.~Briskin, 
  proceedings of
  the workshop ``Future Physics at HERA'', DESY, Hamburg, 1996, 
  and references therein.

\bibitem{ALAN}
 A.~R.~White, ``The Hard Gluon Component of the QCD Pomeron'',
hep-ph/9609282.

\bibitem{ReggeFits}
A.~Kaidalov, Phys.~Rep.~{\bf 50}(1979) 157. \\
G.~Alberi, G.~Goggi, Phys.~Rep.~{\bf 74} (1981) 1. \\
K.~Goulianos, Phys.~Rep.~{\bf 101} (1983) 169. \\
N.~Zotov, V.~Tsarev, Sov.~Phys.~Usp.~{\bf 31}(1988) 119.

\bibitem{XPDETA} E.~L.~Berger et al.~, Nucl.~Phys.~{\bf B286}(1987) 704.


\bibitem{Bjorken} J.~D.~Bjorken, ``Hadronic Final States in Deep--Inelastic
Processes'', in ``Current Induced Reactions: International Summer Institute
in Theoretical Particle Physics in Hamburg 1975'', ed.~J.~G.~Kr\"{o}ner,
G.~Kramer, and D.~Schildknecht, Lecture Notes in Physics, 
Springer Verlag 1976.

\bibitem{Mueller} A.~M\"{u}ller, Phys.~Rev.~{\bf D2}~(1970) 2963.

\bibitem{VMD}
J.\ J.\ Sakurai,
Phys.\ Rev.\ Lett.\ {\bf 22} (1969) 981\\
J.\ J.\ Sakurai and D.\ Schildknecht,
Phys.\ Lett.\ {\bf 40B} (1972) 121, \\
T.\ H.\ Bauer, R.\ D.\ Spital, D.\ R.\ Yennie, F.\ M.\ Pipkin,
Rev.\ Mod.\ Phys.\ {\bf 50} (1978) 261.

\bibitem{QCD93} K.~Golec--Biernat and J.~Kwiecinski, Phys.~Lett.~{\bf B353}.

\bibitem{H1PARIS} J.~Dainton and J.~Phillips, Proceedings of the 
Workshop on Deep-Inelastic Scattering and QCD, Paris, France, April 
24-28 1995.

\bibitem{SPACAL} H1 SpaCal Group, T.~Nicholls et al., 
DESY preprint 95-165 (1995) and DESY preprint 96-013 (1996).

\bibitem{DGLAP}
Yu.\ L.\ Dokshitzer, JETP {\bf 46} (1977) 641.  \\
V.\ N.\ Gribov and L.\ N.\ Lipatov, Sov.\ Journ.\ Nucl.\ Phys.\ {\bf 15} (1972) 78.\\
G.\ Altarelli and G.\ Parisi, Nucl.\ Phys.\ {\bf B126} 1977 298.

\bibitem{GS} T.~Gerhmann and W.~Stirling, Z.~Phys~{\bf C70}, 89 (1996).

\bibitem{ZEUSFPS} 
E.~Gallo, M.~Grothe, C.~Peroni, J.~Rahn, R.~Sacchi, A.~Solano,
  proceedings of
  the workshop ``Future Physics at HERA'', DESY, Hamburg, 1996.

\bibitem{ZEUS_LPS1}
G.~Wolf, Proceedings of the Topical Conference on Hard Diffractive 
Processes, Eilat, Israel, February 1996.

\bibitem{ZEUS_LPS2} 
E.~Barberis, Proceedings of the International Workshop on Deep--Inelastic 
Scattering and Related Phenomena, Rome, Italy, April 1996.


\bibitem{BUCHMUELLER}
W.~Buchm\"{u}ller and A.~Hebecker, Phys.~Lett.~{\bf B355}, 573 (1995), \\
W.~Buchm\"{u}ller and A.~Hebecker, hep-ph/9512329 (1995), \\
W.~Buchm\"{u}ller, M.~F.~McDermott, and A.~Hebecker, 
DESY preprint 96-126 (1996).

\bibitem{SCI}
A.~Edin, J.~Rathsman and G.~Ingelman, Phys.~Lett.~{\bf B366}, 371 (1996), \\
A.~Edin, J.~Rathsman and G.~Ingelman, DESY-96-060 (1996).

\bibitem{VDM} 
ZEUS Collaboration, M.~Derrick et al.~, Physics Letters B 356 (1995) 601-616, \\
H1 Collaboration, S.~Aid et al.~, Nucl.Phys. {\bf B463} (1996) 3. 

\bibitem{FLHWS} 
L.~Bauerdick, A.~Glazov, M.~Klein,   proceedings of
  the workshop ``Future Physics at HERA'', DESY, Hamburg, 1996.

\bibitem{GUNNAR}
G.\ Ingelman and K.\ Janson--Prytz,
proceedings of the workshop ``Physics at HERA", p. 233,
October 1991, ed. W.\ Buchm\"uller and G.\ Ingelman, \\
G.\ Ingelman and K.\ Prytz,
Zeit.\ Phys.\ {\bf C58} (1993) 285.

\bibitem{H1F2} H1 Collaboration, S.~Aid et al.~, 
Nucl.~Phys.~{\bf B470} (1996) 3.

\bibitem{ZEUSF2} ZEUS Collaboration, M.~Derrick et al.~, 
DESY 96-076 (1996).

\bibitem{RAPGAP}
H.\ Jung, Comp.\ Phys.\ Comm.{\bf 86} (1995) 147, \\
H.\ Jung, RAPGAP 2.2 Program Manual, to appear in Comp.\ Phys.\ Comm.
\bibitem{pdg96}
R.\ M.\ Barnett et al.\ (Particle Data Group):
Phys.\ Rev.\ D54~1~(1996).



\end{thebibliography}
\end{document}